\documentclass[10pt,onecolumn]{article}

\usepackage{amsmath}
\usepackage{amssymb}
\usepackage{graphicx}
\usepackage{morefloats}
\usepackage{color}
\usepackage{cite}

\title{\textbf{Dynamics to equilibrium in Network Games: \\ individual behavior and global response}}
\author{Giulio Cimini$^{1,2}$, Claudio Castellano$^2$ and Angel S\'anchez$^{1,3}$\\ \ \\
$^1$ Grupo Interdisciplinar de Sistemas Complejos (GISC)\\
Departamento de Matem\'aticas, Universidad Carlos III de Madrid,\\ 28911 Legan\'es, Madrid, Spain\\[1mm]
$^2$ Istituto dei Sistemi Complessi, (ISC-CNR) and Dipartimento di Fisica,\\
"Sapienza" Universit\'a di Roma, 00185 Roma, Italia\\[1mm]
$^3$ Instituto de Biocomputaci\'on y F\'\i sica de Sistemas Complejos (BIFI),\\ Universidad de Zaragoza, 50018 Zaragoza, Spain }
\date{This version: \today}

\begin{document}
\maketitle

\begin{abstract}
Various social contexts ranging from public goods provision to information collection
can be depicted as games of strategic interactions, 
where a player's well-being depends on her own action as well as on the
actions taken by her neighbors. 
Whereas much attention has been devoted to the identification and 
characterization of Bayes-Nash equilibria of such games, in this work 
we look at strategic
interactions from an evolutionary perspective. Starting from a recent
mean-field analysis of the evolutionary dynamics in these games, 
here we present results of numerical simulations designed to find out whether
Nash equilibria are accessible by adaptation of players' strategies, 
and in general to find the attractors of the evolution. 
Simulations allow us to go beyond a global characterization of the 
cooperativeness of the equilibria 
and probe into the individual behavior.
We find that when players imitate each other, the evolution does not reach
Nash equilibria and, worse, leads to very unfavorable states in terms of
welfare. 
On the contrary, when players update their behavior rationally, 
they self-organize into a rich variety of Nash equilibria, 
where individual behavior and payoffs are shaped by the nature of the game, 
the structure of the social network and the players' position within 
the topology. 
Our results allow us to assess the validity of the mean-field approaches 
and also show qualitative agreement with theoretical predictions for
equilibria in the context of one-shot games under incomplete information.
\end{abstract}

\section{Introduction}

In a range of social and economic interactions (public goods
provision, job search, political alliances, trade, friendships,
information collection, and so on) an individual's welfare depends
both on her own actions and on the actions taken by her interacting
partners.  Much effort has been devoted to understand how the pattern
of social connections shapes the choices that individuals make and the
payoffs that they can
earn\cite{Katz1955,Coleman1966,Granovetter1994,Foster1995,Glaeser1996,Topa2001,Conley2010}
(see \cite{VegaRedondo2007,Goyal2007,Jackson2008} for an overview of
the field).  On the theoretical side, the traditional approach of
identifying the Bayes-Nash equilibria in one-shot network games 
is problematic due to the existence of multiple equilibria (even for very
small network sizes), which imply a huge range of possible outcomes.
Nevertheless, it has been shown that the problem of multiplicity can
be resolved sometimes by the introduction of incomplete information
\cite{Carlsson1993,Morris2003}---which in the context of network games
means having only a local knowledge of the network: a player is aware
of the number of her connections (her degree) but not of the degrees
of the others.  The key point is that when players have limited
information about the network they are unable to condition their
behavior on its fine details, and this leads to a significant
simplification and sharpening of equilibrium predictions.  A general
framework for the study of multiplayer games on networks under
incomplete information has been recently developed
\cite{Bramoulle2007,Galeotti2010}; in particular, Galeotti et
al. \cite{Galeotti2010} proved the existence of Bayes-Nash equilibria
involving strategies that are monotonic with respect to players'
degrees and symmetric, \emph{i.e.}, with all players of the same degree choosing
the same strategy.

Here, instead of focusing on equilibrium strategies played once and
for all, we consider iterated network games within an evolutionary framework
\cite{MaynardSmith1973}: players' own actions are described by a
strategy which is subject to an evolutionary process
\cite{Hofbauer1998,Gintis2009}. It is important to note that this is
different from the stochastic stability approach presented for this
type of games in \cite{boncinelli:2012}, where the focus is on the
different time regimes of the evolution.
In fact, our framework represents a way to select among the large
multiplicity of equilibria that exist in network games in a spirit
very close to that of
biological evolution.  In particular, we consider two main mechanisms
for players to adapt their strategy: imitation and rational
deduction. Moreover, we consider two representative social network
structures (Erd\"{o}s-R\'{e}nyi random graphs \cite{Erdos1960} and
scale-free networks \cite{Barabasi1999}) and, as in
\cite{Galeotti2010}, two canonical types of interaction: the best-shot
game and the coordination game, as representatives for strategic
substitutes and strategic complements, respectively
\cite{Bulow1985}. We run a numerical simulation program in order to
understand which Nash equilibria are dynamically accessible (by
adaptation of players' strategies), and in general to identify the
strategy configurations that are the attractors of such
dynamics---either Nash equilibria or, possibly, other types of
stationary states. As a reference, we consider the analytical
results we obtained in \cite{us}, where we studied this issue by
means of two mean field approaches, the usual, homogeneous one, and
its heterogeneous version \cite{PastorSatorras2001}. As shown
below, the numerical simulations allow us to assess the accuracy
of the pictures provided by the two approaches and, furthermore, to
gain more insight on the individual behavior of the different types of players.
Additionally, we compare our approach with the theoretical framework
of Galeotti et al.~\cite{Galeotti2010} for one-shot games under incomplete
information.  We show that the Nash equilibria predicted by their theory 
generally do not coincide with the ones obtained by evolutionary dynamics;
nevertheless, these equilibria possess qualitatively the same features 
of those resulting from evolution, in terms of players'
strategies with respect to their neighborhood structure.

Our work is organized as follows. We begin by introducing in Section 2
the games we will be considering, followed by a presentation of the
two evolutionary dynamics in Section 3. Section 4, the main part
of the paper, presents the results of numerical simulations, which,
for reference, are accompanied by a brief review of the mean field
results in \cite{us}.  Finally, Section 5 concludes the paper with a
summary of the results and a discussion of how they compare to the
different theoretical frameworks.

\section{Games and Equilibria}

This section presents two simple games played on networks, reflecting
strategic substitutes and strategic complements, respectively
\cite{Bulow1985}.  These two cases represent alternative scenarios on
how a player's payoff is affected by the actions of others, covering
many of the game-theoretic applications studied in the economic
literature.  In particular, strategic substitutes encompass many
scenarios that allow for free riding or have a public good structure
of play, whereas strategic complements arise whenever the benefit that
an individual obtains from undertaking a given behavior is greater as
more of her partners do the same.

Below, we first recall the framework by Galeotti et al. \cite{Galeotti2010}
under the assumption of incomplete information, which is natural
in many circumstances.
Consider a society of $n$ agents, placed on
the nodes of a social network. The links between agents reflect social
interactions, and connected agents are said to be \emph{neighbors}.
Every individual must choose independently an action in $X=\{0,1\}$,
where action 1 may be interpreted as \emph{cooperating} and action 0
as not doing so---or \emph{defecting}.  To define the payoffs, let
$x_i$ be the action chosen by agent $i$, $N_i$ the set of $i$'s
neighbors, $x_{N_i}=\sum_{j\in N_i}x_j$ the aggregate action in $N_i$,
and $y_i=x_i+x_{N_i}$.  There is a cost $c$, where $0<c<1$, for
choosing action $1$, while action $0$ bears no cost.

\subsubsection*{Strategic Substitutes: Best-shot game}
The payoff function takes the form
\begin{equation}\label{eq.SS}
\pi_i=\Theta_H(y_i-1)-c\,x_i,
\end{equation}
where $\Theta_H(\cdot)$ is the Heaviside step function $\Theta_H(x)=1$ if
$x\ge0$ and $\Theta_H(x)=0$ otherwise.  Strategic substitutes thus
represent an anti-coordination game: a player would prefer that
someone of her neighbors takes action 1 (rather than taking the action
herself), but she would be willing to take action 1 if nobody in the
neighborhood does.  In general, a context in which players have
complete information on the social network allows for a very rich set
of Nash equilibria of the game \cite{Bramoulle2007,Galeotti2010},
where the relation between network connections, equilibrium actions
and payoffs may exhibit very different patterns (even when all agents
of the same degree choose the same actions).  Things change, however,
by relaxing the assumption of complete information---as shown in
\cite{Galeotti2010}.  The assumption is that each player
knows her own degree $k'$ and the probability distribution $P(k|k')$ 
of the degree $k$ of her neighbors, which for uncorrelated networks
(degrees of neighboring nodes are independent) reads
\begin{equation}\label{eq.PQ}
P(k|k')=kP(k)/\bar{k},
\end{equation}
where $P(k)$ is the degree distribution of the network and 
$\bar{k} = \sum_k k P(k)$ the average connectivity.
Under these conditions, Galeotti et al.~\cite{Galeotti2010} study the game
within the framework of Bayesian games
and show that a player's (pure) strategy 
$\sigma \in X$ depends only on the degree $k$ of the player.
If an agent of degree $k$ chooses action 1 in equilibrium, 
it must be because she does not expect that any of her neighbors will choose action 1. 
Therefore, in an uncorrelated network, 
an agent of degree $k-1$ faces a lower likelihood of an arbitrary neighbor 
choosing the action 1, and would be best responding with action 1 as well. 
In particular\cite{Galeotti2010}, any Nash equilibrium is characterized 
by a threshold $\tau$, the smallest integer for which
\begin{equation}\label{eq.SS_tau}
1-\left[1-\sum_{k=1}^\tau \frac{kP(k)}{\bar{k}} \right]^\tau\ge 1-c,
\end{equation}
and an equilibrium $\sigma$ must satisfy $\sigma(k)=1$ for
all $k<\tau$, $\sigma(k)=0$ for all $k>\tau$ and
$\sigma(\tau)\in\{0,1\}$ (\emph{i.e.} $\sigma(k)$ is non-increasing).

\subsubsection*{Strategic Complements: Coordination game}
The payoff function here takes the form
\begin{equation}\label{eq.SC}
\pi_i=(\alpha x_{N_i}-c)\,x_i,
\end{equation}
where $0<\alpha<c$. Strategic complements thus represent a
coordination game.  As for substitutes, in the case of complete
information there are generally many equilibria (including the case
$x_i=0$ $\forall i$, which represents full defection).  Also here,
by making the assumption that each player is only informed of her own
degree and has independent beliefs on the degrees of neighbors, it is
possible to find much more definite predictions with regard to
equilibrium behavior \cite{Galeotti2010}. 
In particular, independence of neighbor degrees implies
that the probability that a random neighbor chooses the action 1
cannot depend on one's own degree 
(a player's neighbors do not know her degree and, consequently, 
they cannot know whether or not it will be convenient for her to chose action 1), 
and the expectation of the sum of
actions $x_{N_i}$ of any agent $i$ with $|N_i|=k$ neighbors is
increasing in $k$. The structure of payoffs then assures that if a
degree $k$ agent is choosing the action 1 in equilibrium, any agent of
degree greater than $k$ must be best responding with the action 1 as well, 
as she should have as many cooperating neighbors as an agent of degree $k-1$.  
Therefore every equilibrium is characterized by an integer
threshold $\tau$~\cite{Galeotti2010} such that
\begin{equation}\label{eq.SC_tau}
\alpha(\tau-1)\sum_{k=\tau}^{n-1} \frac{kP(k)}{\bar{k}} <c\qquad\mbox{and}
\qquad\alpha\tau\sum_{k=\tau}^{n-1} \frac{kP(k)}{\bar{k}} \ge c.
\end{equation}
The equilibrium satisfies $\sigma(k)=0$
for all $k<\tau$, $\sigma(k)=1$ for all $k>\tau$ and
$\sigma(\tau)\in\{0,1\}$ (in particular, $\sigma(k)$ is
non-decreasing).
\newline\newline

Summing up, for equilibrium strategies played once and for all under
complete network information there is no systematic relation between
social networks and individual behavior and payoffs. 
By contrast, under
incomplete network information, both in games of strategic substitutes
and of strategic complements, there is a clear cut relation between
networks and individual behavior, as decisions can in general 
be inferred from degrees.

\section{Evolutionary Dynamics}

Whereas the previous section dealt with Bayes-Nash equilibria of
one-shot strategic games under the settings of complete and incomplete
information, here we move to our contribution, namely an evolutionary scenario in which agents
play multiple instances of the game, and can adapt their strategy
hoping to increase their payoffs.  We resort to an evolutionary
game-theoretical approach in which agents do not make strategic
considerations about the network's global structure and their position within
it; instead, they base their decision on their own 
actions/payoffs and on those of neighbors, as observed in the past.
Our goal is to study the resulting dynamics of the system as well
as its equilibrium configurations, and check whether the system can reach
a Nash equilibrium, or what is the nature of the steady state if not Nash.

The evolutionary system can be described as follows. Starting with a
fraction $\rho_0$ of players randomly chosen to undertake action
$x=1$, at each round $t$ of the game: players collect their payoff
$\pi^{(t)}$---given by Eq.(\ref{eq.SS}) and Eq.(\ref{eq.SC})
respectively for strategic substitutes and complements; then a
fraction $q$ of players update their strategy.
We consider two different mechanisms for strategy updating.

\emph{Proportional Imitation} (PI) \cite{Helbing1992} --- It
represents a rule of imitative nature in which player $i$ may copy the
strategy of a selected counterpart $j$, which is chosen randomly among
the $N_i$ neighbors of $i$. The probability that $i$ copies $j$'s
strategy depends on the difference between the payoffs that they
obtained in the previous round of the game:
\begin{equation}\label{eq.PI}
\mathcal P\left\{x_j^{(t)}\rightarrow x_i^{(t+1)}\right\}=\begin{cases}
	(\pi_j^{(t)}-\pi_i^{(t)})/\Phi&\mbox{if $\pi_j^{(t)}>\pi_i^{(t)}$},\\
	0&\mbox{otherwise},
	\end{cases}
\end{equation}
where $\Phi$ is a normalization constant that ensures $\mathcal
P\{\cdot\}\in[0,1]$.  Note that because of the imitation mechanism of
PI, the configurations $x_i=1$ $\forall i$ and $x_i=0$ $\forall i$ are
absorbing states---the system cannot escape from them.

\emph{Best Response} (BR) \cite{Matsui1992,Blume1993} --- Here players
are fully rational and choose their strategy in order to maximize
their payoff, given what their neighbors did in the last round.  This
means that each player $i$, given $x_{N_i}^{(t)}$, computes the
payoffs that she would obtain by choosing action 1 (cooperating) or 0
(defecting) at time $t$, respectively $\tilde{\pi}_C^{(t)}$ and
$\tilde{\pi}_D^{(t)}$.  Then
\begin{equation}\label{eq.BR}
x_i^{(t+1)}=\begin{cases}
	1&\mbox{if $\tilde{\pi}_C^{(t)}>\tilde{\pi}_D^{(t)}$},\\
	0&\mbox{if $\tilde{\pi}_C^{(t)}<\tilde{\pi}_D^{(t)}$},\\
	x_i^{(t)}&\mbox{if $\tilde{\pi}_C^{(t)}=\tilde{\pi}_D^{(t)}$},
	\end{cases}
\end{equation}

We use PI because it is equivalent, for a well-mixed population,
to the well-known replicator dynamics \cite{Gintis2009}, and BR because it
is widely used in the economic literature.  We then study how the
system evolves starting from the initial random distribution of
strategies. In particular, we want to check whether the system can
reach a Nash equilibrium (a state where no player can increase her
payoff by unilaterally changing her strategy).  Note, however, that
whereas a Nash equilibrium is stable by definition under BR dynamics,
with PI this is not necessarily true: players can change action by
copying better-performing neighbors, even if such change deteriorates
their payoffs.  Hence in the latter case a potential stationary state
of the system does not necessarily correspond to a Nash equilibrium 
(unless the network is a complete graph \cite{Gintis2009}).

In what follows we present analytical and numerical results for the
system described above.  For simulations, without loss of generality we set
$\rho_0=1/2$, $c=1/2$ and $q=1/10$---but our results are valid for all
values of these parameters in $(0,1)$.\footnote{Note that we have 
to impose the condition $q<1$ to avoid possible trapping into period-2 loops, 
where all players change action simultaneously at each round of the game, 
that can arise in anti-coordination under BR.} 
We will show averages over $\mathcal{N}=20$ independent realizations 
of the system, an amount that we found enough to tame single-realization 
fluctuations (that are however shown with error bars, when visible). 
Note that in numerical simulations, when the system arrives to a Nash equilibrium, 
we change by hand the strategy of a random player in order to continue exploring
other strategy configurations.
\newline\newline

As stated in the introduction, we consider two representative kinds of
population structure.
Erd\"{o}s-R\'{e}nyi (ER) random graphs~\cite{Erdos1960} are built by
considering a collection of $n$ nodes and adding a link between
each pair of nodes with probability $p$. The resulting networks
are homogeneous, with degree distribution decaying exponentially for
large degree $k$ and the average degree is $\bar{k}=(n-1)\,p=m$.
Scale-free (SF) random networks are generated using a configuration
model~\cite{Newman2003} with a constraint $k_{max}<\sqrt{n}$, 
which gives rise to uncorrelated random networks \cite{Catanzaro2005}
with degree distribution $P(k)\propto k^{-\gamma}$.
The average degree depends on the network size $n$ and converges to a
finite value as $n$ diverges
\footnote{In what follows
  we set $\gamma=2.5$ and $k_{min}=3$, and consider various sizes:
  $n=1\,000$ ($\bar{k}=6.415$), $n=10\,000$ ($\bar{k}=7.48$),
  $n=100\,000$ ($\bar{k}=8.13$), $n=1\,000\,000$ ($\bar{k}=8.51$).
  Note that for $n\rightarrow\infty$ it is $m:=\bar{k}_\infty=9$.}
$$\bar{k}=\frac{\gamma-1}{\gamma-2}\left[\frac{\sqrt{n}^{2-\gamma}-k_{min}^{2-\gamma}}{\sqrt{n}^{1-\gamma}-k_{min}^{1-\gamma}}\right]
\xrightarrow[n\rightarrow\infty]{}\frac{\gamma-1}{\gamma-2}\,k_{min}=:m.$$.

\section{Results}

We now present the results of the numerical simulations. We recall
that in Ref.~\cite{us} we analyzed the evolutionary dynamics for the two
games, under the same two evolutionary rules, by resorting to two
analytical approaches, the homogeneous mean-field (MF) and the heterogeneous 
mean-field (HMF). While we refer to Ref.~\cite{us} for the complete
analysis and detailed discussion, we 
include brief summaries of the analytical predictions in the following,
as tools to understand the simulation results.
Note that the MF approach, which should 
work best for homogeneous random graphs, deals with the
aggregate density of cooperators $\rho(t)$, whereas, the HMF, which is
more appropriate for heterogeneous networks, involves the computation of
the quantity $\Theta(t)=\sum_{k}kP(k)\rho_{k}(t)/\bar{k}$
(\emph{i.e.}, the average of the probability $\rho_k$ that a node of 
degree $k$ cooperates, weighted by the relative degree).

\subsection{Best-shot game}

\subsubsection*{Proportional Imitation}

\begin{figure}[h!]
\centering
\includegraphics[width=0.5\textwidth]{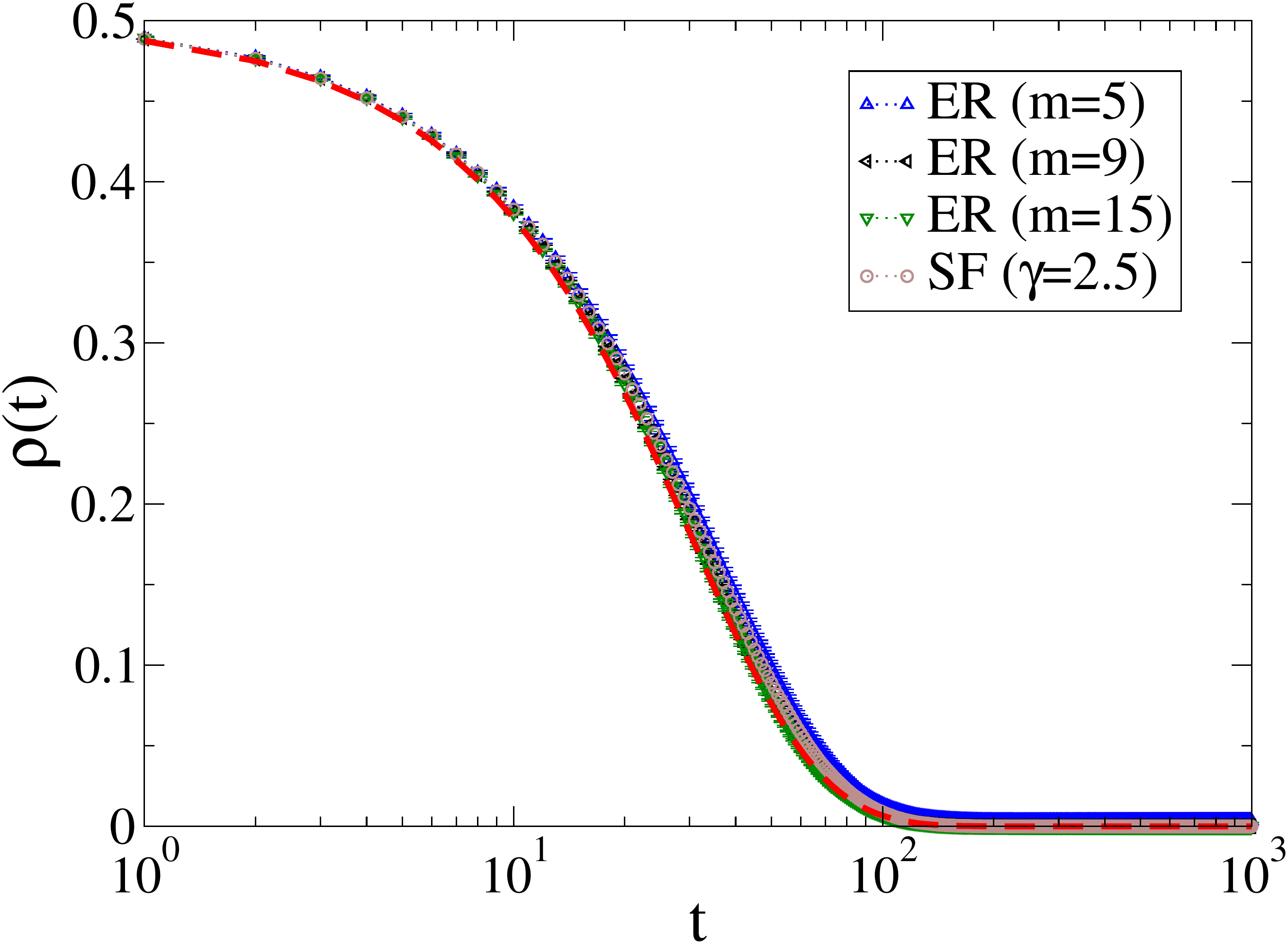}
\caption{Best-shot games with Proportional Imitation: 
average $\rho(t)$ for Erd\"{o}s-R\'{e}nyi and scale-free graphs, with $n=10^4$ 
(but results are independent on the specific value of $n$). 
The red dashed curve is the MF result---Eq.(\ref{eq.SS_PI_sol}).}\label{fig.SS_PI}
\end{figure}

In this case, the MF prediction is that
\begin{equation}\label{eq.SS_PI_sol}
\rho(t)=[1+(\rho_0^{-1}-1)\exp(cqt)]^{-1},\qquad 0<\rho_0<1.
\end{equation}
Hence the population converges to the state with no
cooperators $\rho=0$ (full defection), unless the initial state is
$\rho_0=1$ (full cooperation).
We see from Figure
\ref{fig.SS_PI} that, also in simulations, the system always goes towards 
the absorbing state $x_i=0$ $\forall i$ $\Rightarrow$ $\rho\equiv0$
and the full dynamical evolution is in excellent agreement with 
the MF theory.
The convergence toward full defection occurs because a defector 
cannot copy a neighboring cooperator 
(who has lower payoff by construction), whereas a cooperator will
eventually copy one of her neighboring defectors (who has higher payoff). 
However, full defection is not a Nash equilibrium as any player surrounded by
defectors would do better by cooperating.

In the case of scale-free networks, the behavior is again remarkably
well described by the simple MF approach, which coincides with the 
HMF as $\rho_k(t=0)$ does not depend on $k$.

\subsubsection*{Best Response}

For the best-shot game with BR dynamics, the MF predicts that the
final state is, for any initial condition, a mixed state 
$\rho=\rho_s$, where $\rho_s\in(0,1)$ is the solution of the equation:
\begin{equation}\label{eq.SS_BR_2}
\rho_s=e^{-m\rho_s}.
\end{equation}
The HMF approach leads to an analogous conclusion: the final state is
$\Theta_s$, which is the solution of the equation
\begin{equation}\label{eq.SS_BR_h_3}
\Theta_s=\sum_k(1-\Theta_s)^kkP(k)/\bar{k}.
\end{equation}
Since BR dynamics is guaranteed to lead to Nash equilibria,
$\rho_s$ and $\Theta_s$ are the attractors of the dynamics, their
values depending only on the average degree of the network (in
particular, they decrease for increasing network connectivity) but
not on $\rho_0$, $c$ or $q$.
\newline

\begin{figure}[t!]
\centering
\includegraphics[width=\textwidth]{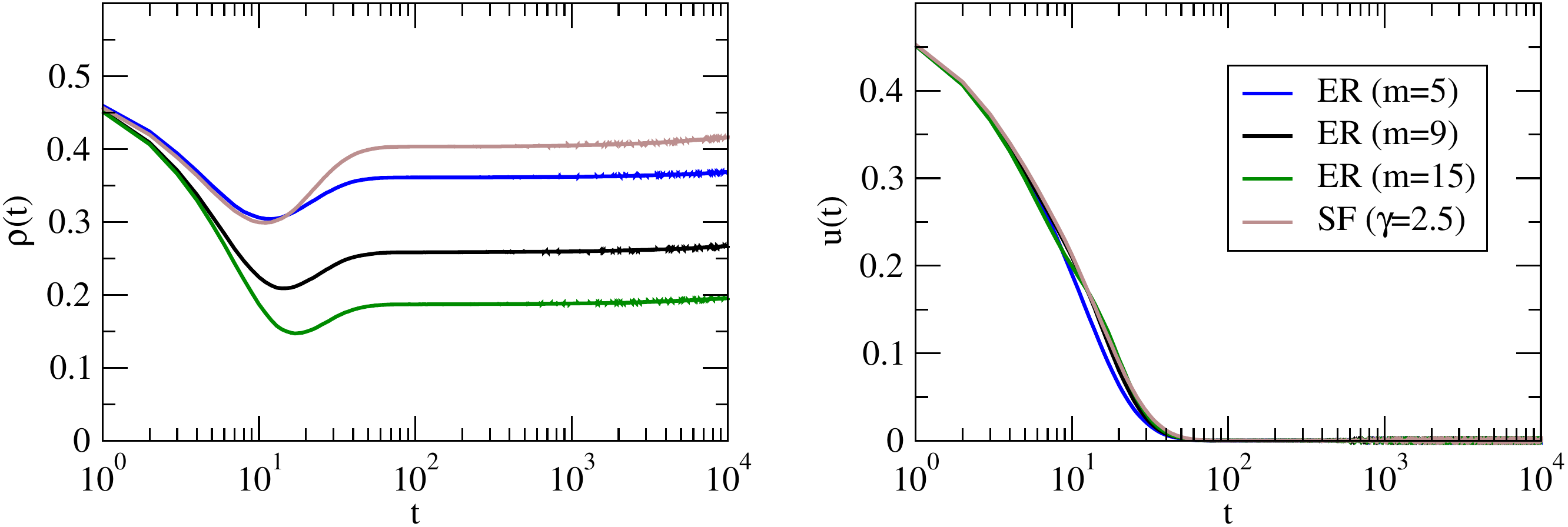}
\caption{Best-shot games with Best Response: averages of $\rho(t)$ and
  fraction $u(t)$ of unsatisfied players (\emph{i.e.}, who would obtain higher payoff 
  by unilaterally switching action, and with $u=0$
  indicating the falling of the system into a Nash equilibrium) 
  for Erd\"{o}s-R\'{e}nyi and scale-free graphs, with $n=10^4$ 
  (again, results are independent on the specific value of $n$).}\label{fig.SS_BR}
\end{figure}

\begin{figure}[h!]
\centering
\includegraphics[width=0.5\textwidth]{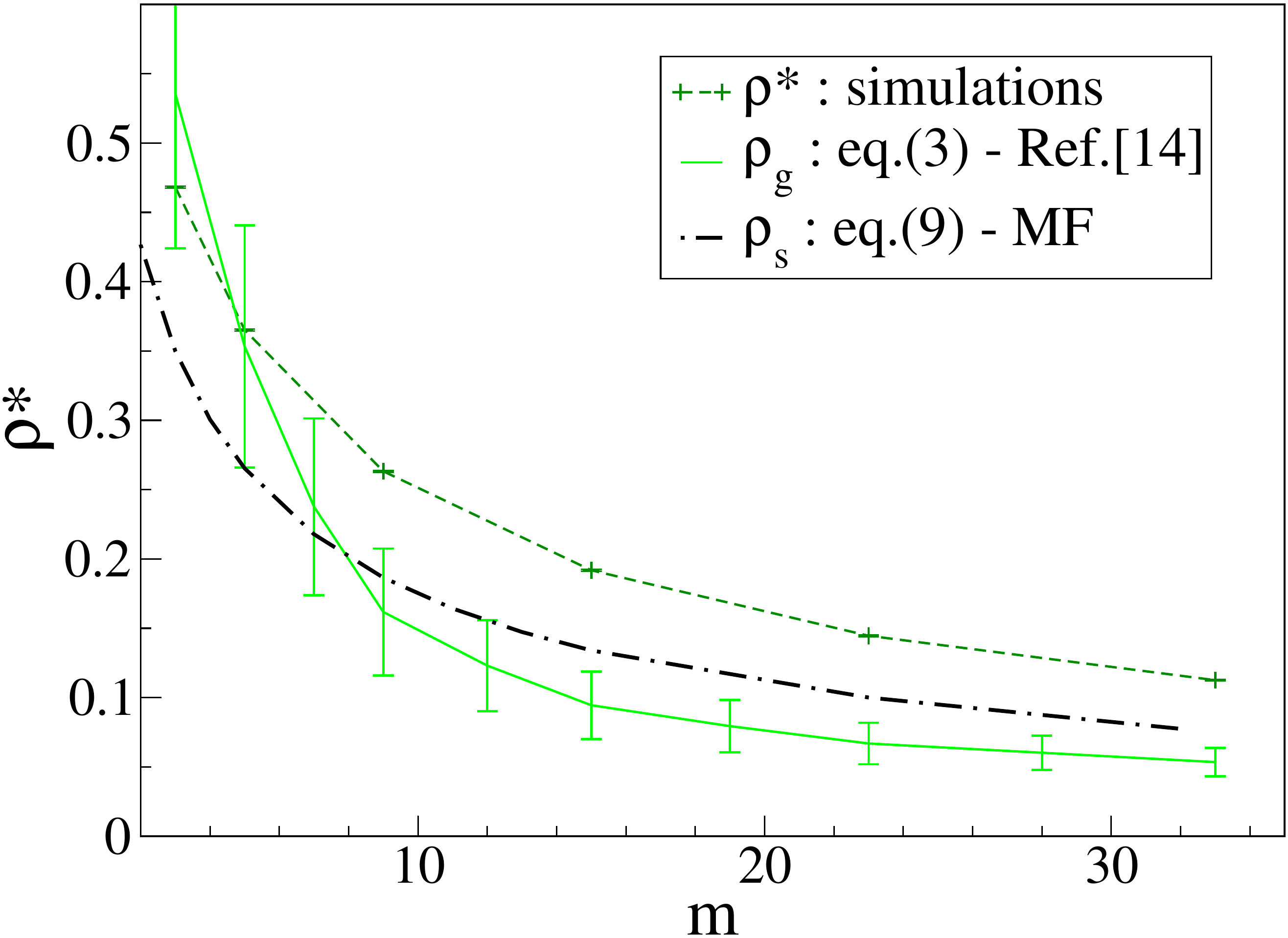}
\caption{Best-shot games with Best Response on Erd\"{o}s-R\'{e}nyi random graphs: 
  average $\rho^*$ at Nash equilibria vs $m$ from simulations ($n=10^4$), 
  theoretical prediction from Eq.(\ref{eq.SS_tau}), and MF Eq.(\ref{eq.SS_BR_2}).}\label{fig.SS_BR_eq}
\end{figure}

\begin{figure}[h!]
\centering
\includegraphics[width=0.85\textwidth]{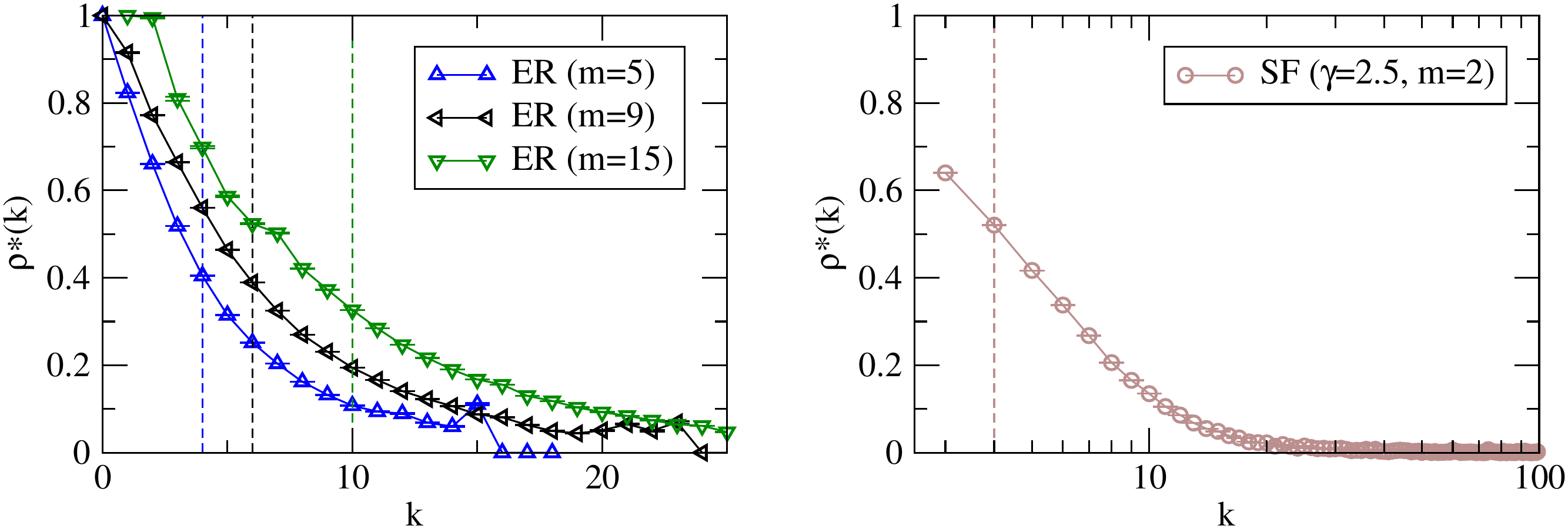}
\caption{
Best-shot games with Best Response: average $\rho^*(k)$ at
  Nash equilibria for nodes with different degrees $k$. The vertical
  dashed lines identify the thresholds $\tau$ from
  Eq.(\ref{eq.SS_tau})---which basically depend on
  $m$.}\label{fig.SS_BR_degree}
\end{figure}

Numerical simulations indeed confirm such a picture: the dynamics finds
a rich variety of Nash equilibria with intermediate cooperation level
$\rho^*$ (Figure \ref{fig.SS_BR}); moreover, $\rho^*$ decreases with
increasing network connectivity (Figure~\ref{fig.SS_BR_eq})---in
agreement with both Eq.~(\ref{eq.SS_tau}) and Eq.~(\ref{eq.SS_BR_2}).
The key observation to understand the features of such equilibria is
that, in best-shot games under BR dynamics, a player switches to
defection as soon as one of her neighbors is a cooperator, and this
happens with higher probability when the player has many social ties
(see also Figure~\ref{fig.SS_BR_degree}).
The higher values of $\rho^*$ in scale-free networks than 
in Erd\"{o}s-R\'{e}nyi graphs (of the same link density) can be explained by the
presence, in the first case, of more low-degree nodes---who
preferentially cooperate as they have few neighbors to exploit. 
However, low-degree nodes weight less in $\Theta$, so that we
find a $\Theta^*$ for random scale-free networks similar to the
$\rho^*$ of Erd\"{o}s-R\'{e}nyi random graphs (in agreement with the
mean field predictions).
\footnote{This is due to nodes with the
  highest degrees---that make the difference in $P(k)$---not cooperating
  in the best-shot game, so that their effects on the system is
  negligible and $\rho_s\simeq\Theta_s$.}  

The Nash equilibria found dynamically can be compared with those
derived under the assumption of incomplete information
by \cite{Galeotti2010}---Eq.(\ref{eq.SS_tau}).  
We first inspect if the cooperation level of our equilibria lies in the range
$\rho_g\in[\rho_{k<\tau},\rho_{k\le\tau}]$, where $\rho_{k<\tau}$ and
$\rho_{k\le\tau}$ are the densities of players with $k<\tau$ and
$k\le\tau$, respectively, and $\tau$ is given by Eq.(\ref{eq.SS_tau}).
Figure \ref{fig.SS_BR_eq} shows the average $\rho^*$ at Nash
equilibria for Erd\"{o}s-R\'{e}nyi random graphs as a function of $m$,
obtained by numerical simulations, by the theoretical framework of
\cite{Galeotti2010}, and by the mean field approach. Clearly, whereas
the trends are similar, we observe that the cooperation
levels of the static equilibria are compatible with simulations
only for small values of $m$, while the mean field approximation works 
better for high average degrees.  
We then check whether $\rho^*(k)$, \emph{i.e.}, the strategy profile 
as a function of the degree $k$ of dynamically-found Nash equilibria, 
is compatible with $\sigma(k)$ from Eq.~(\ref{eq.SS_tau}). Figure 
\ref{fig.SS_BR_degree} shows that, as predicted in \cite{Galeotti2010}, 
$\rho^*(k)$ is generally non-increasing; however, the behavior is not step-like.  
Summing up, the dynamics leads to Nash equilibria which share some
qualitative features but do not coincide with those identified in \cite{Galeotti2010} 
under the assumption of incomplete information.
Note that this does not occur because the configuration space is
not sufficiently explored\footnote{We have tried to manually drive the system 
into one of the equilibria derived by~\cite{Galeotti2010} with
$\sigma(k)=1$ for all $k<\tau$, $\sigma(k)=0$ for all $k>\tau$,
$\sigma(\tau)\in\{0,1\}$. To do so, we set at the beginning of the
evolution a configuration with the action of players with
$k<\tau$ frozen to 1, the action of players with $k>\tau$ frozen to 0,
and the action of players with $k=\tau$ free to evolve according to
BR. Across many realizations, the evolution never converges to a Nash
equilibrium---what happens instead is that all players with $k=\tau$
became satisfied (\emph{i.e.}, they cannot increase their payoffs by unilaterally chaniging action), 
but making unsatisfied the others who cannot change action.}, 
but rather because the two approaches are intrinsically different 
and do not have necessarily to give the same results. 
In fact, the equilibria that are evolutionary selected by our deterministic BR-based dynamics 
are indeed proper Nash, whereas, the framework of \cite{Galeotti2010} is probabilistic 
and thus selects Bayes-Nash equilibria.

\subsection{Coordination game}

For coordination games the picture is richer, as the behavior
of the system depends on the ratio $\alpha/c$.  Without loss of
generality, we leave $c=1/2$ fixed and vary the value of $\alpha$ in
the range $(0,c)$.

\subsubsection*{Proportional imitation}

\begin{figure}[h!]
\begin{minipage}[b]{0.49\textwidth}
\centering
\includegraphics[width=0.975\textwidth]{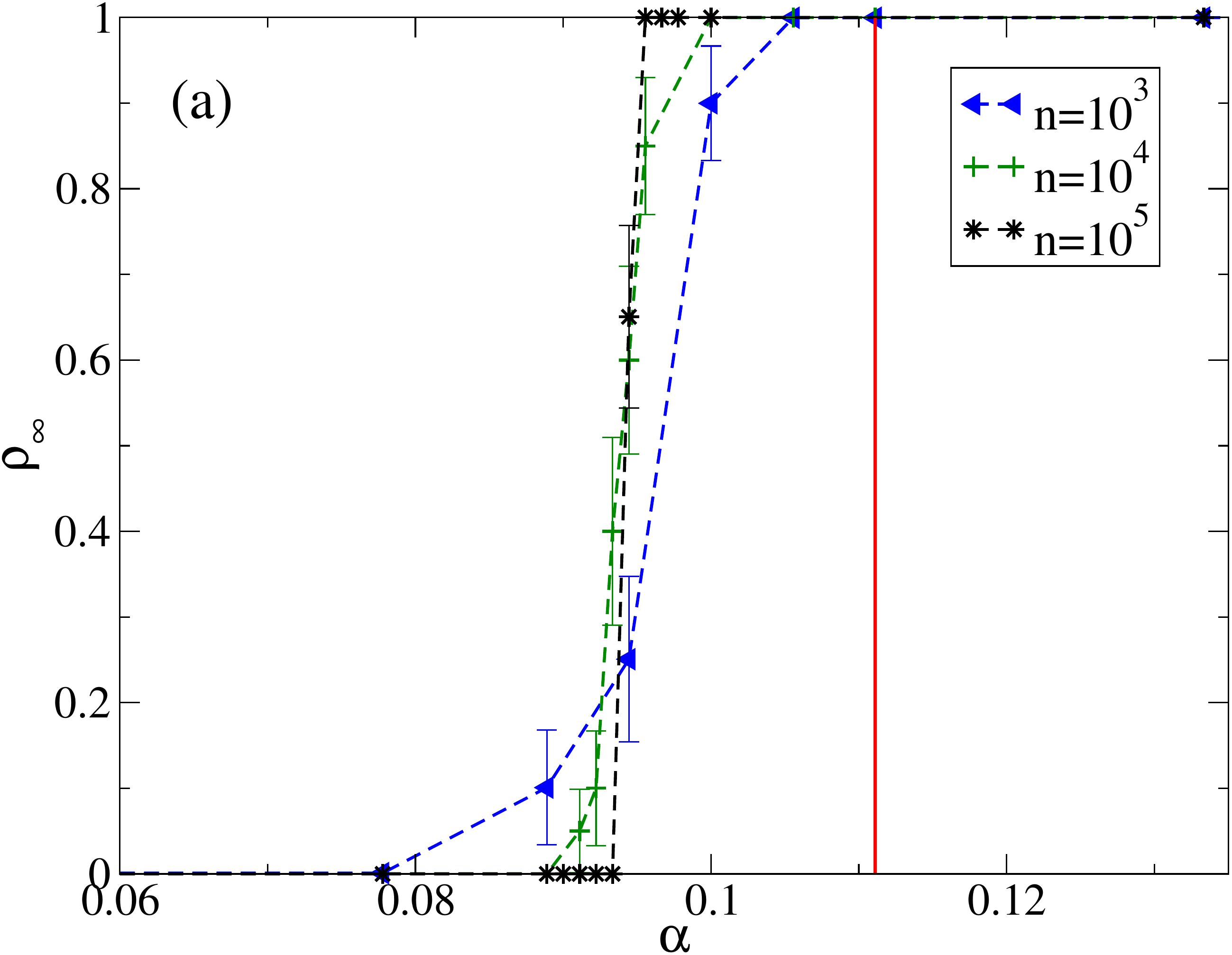}
\end{minipage}
\begin{minipage}[b]{0.49\textwidth}
\centering
\includegraphics[width=0.975\textwidth]{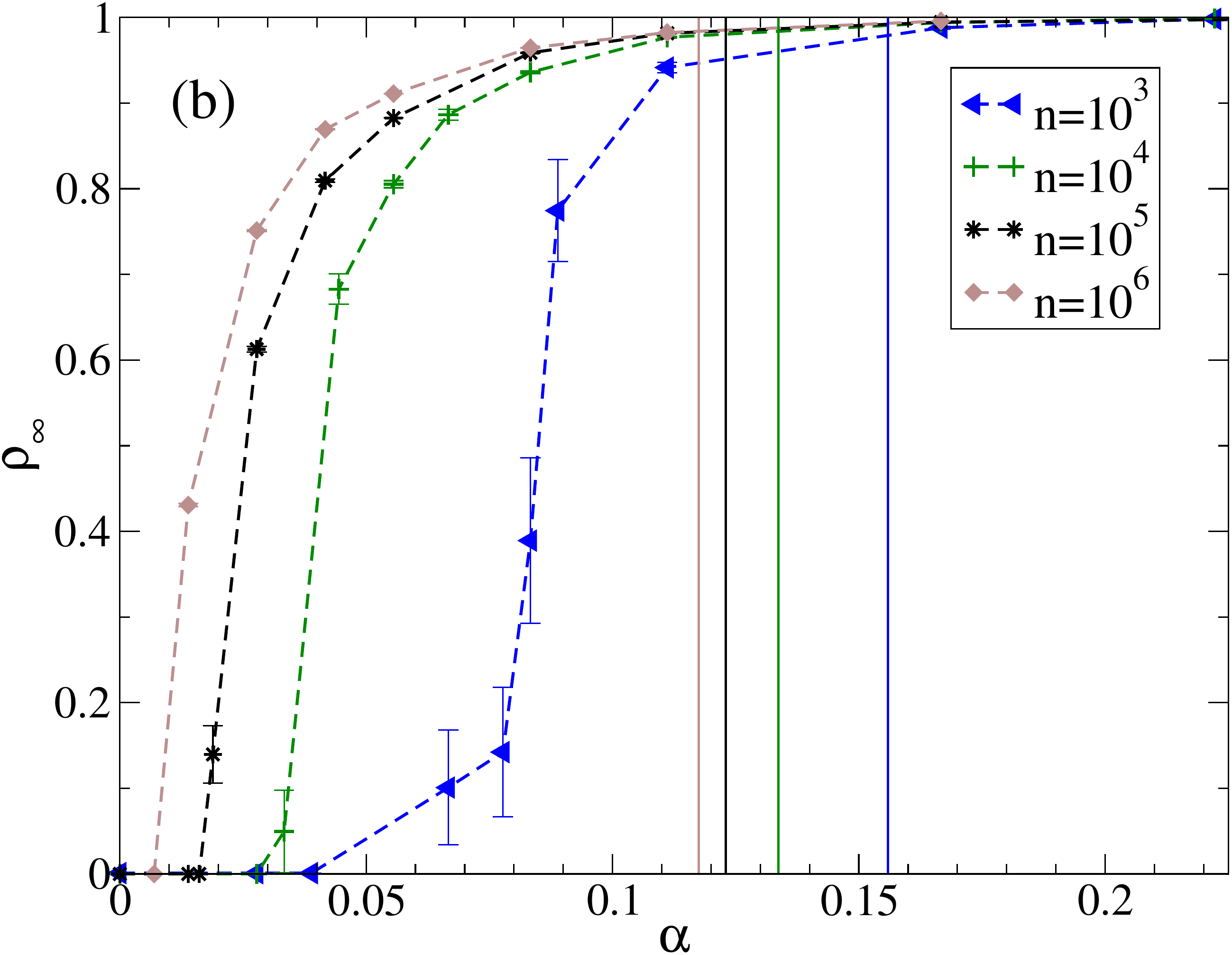}
\end{minipage}
\centering
\includegraphics[width=0.65\textwidth]{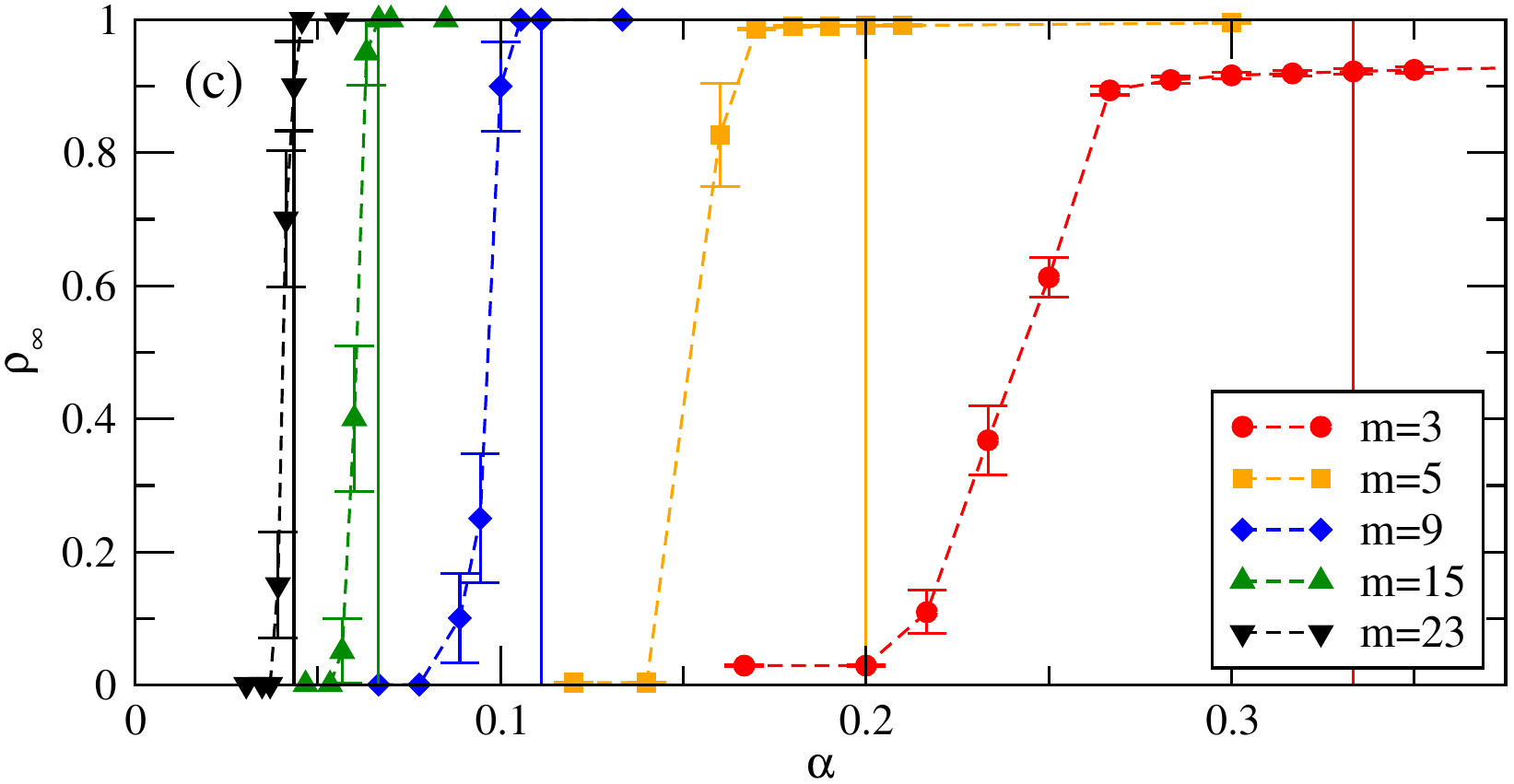}
\caption{Coordination games with Proportional Imitation. 
The vertical solid lines identify the values of $\alpha_c$. 
Erd\"{o}s-R\'{e}nyi random graphs: 
(a) stationary  cooperation levels $\rho_\infty$ vs $\alpha$ for $m=9$ and various $n$;  
(c) stationary cooperation levels $\rho_\infty$ vs $\alpha$ for $n=10^3$ and various $m$. 
  Note that for small values of $m$
  such that $p=m/(n-1)<\ln(n)/n$ the graph is disconnected: isolated
  nodes are bound to their initial action, so that full defection and
  full cooperation are not accessible states; additionally, defective
  behavior spreads easily inside isolated components which are poorly
  connected (also for high values of $\alpha$), so that $\rho_\infty$
  remains far from 1.
Scale-free networks: 
(b) stationary cooperation levels $\rho_\infty$ vs $\alpha$ for $m=9$ and various $n$.}\label{fig.SC_PI}
\end{figure}

The MF theory for the coordination game with PI
predicts the existence of a critical value $\alpha_c = c/(m \rho_0)$,
such that the final state of the dynamics is $\rho=0$ (full defection) 
when $\alpha<\alpha_c$, and $\rho=1$ (full cooperation) when $\alpha>\alpha_c$.
Note, however, that while full defection is always a Nash equilibrium 
for the coordination game, full cooperation becomes a Nash equilibrium 
only when $\alpha>c/k_{min}$,
where $k_{min}$ is the smallest degree in the network---which means
that only networks with $k_{min}>c/\alpha>1$ may feature a fully
cooperative Nash equilibrium.
For Erd\"{o}s-R\'{e}nyi graphs, numerical simulations agree with this 
picture (Figure \ref{fig.SC_PI}), as
a discontinuous transition for $\alpha=\alpha_T$ is observed. 
The value $\alpha_T$ of the transition point found numerically
is smaller than the MF prediction $\alpha_c$;
however $\alpha_T \rightarrow \alpha_c$ as the average degree $m$ grows.
Note that apart from full defection, no other
Nash equilibrium (with intermediate cooperation levels) is found.

The HMF theory for PI dynamics predicts a discontinuous transition
between a stable fully defective equilibrium $\Theta=0$ for 
$\alpha<\alpha_c$ and full cooperation $\Theta=1$ for 
$\alpha>\alpha_c$~\cite{us}.
The critical value is now $\alpha_c=c/\Theta_2$, where
\begin{equation}\label{eq.SS_PI_h_Theta2}
\Theta_2:=\sum_{k}k^2P(k)\rho_{k}/\bar{k}.
\end{equation}
The quantity $\Theta_2$ is related to the second moment of the degree
distribution $\langle k^2\rangle$ and it can be shown to diverge for
networks with $\gamma<3$ as the system size $n$ goes to infinity. 
Therefore, at odds with the case of Erd\"{o}s-R\'{e}nyi 
random graphs, for scale-free networks the threshold $\alpha_c \to 0$ 
as the system size $n$ diverges. 
This vanishing of the transition point is analogous to
what occurs for other processes on scale-free networks, such as 
percolation of epidemic spreading~\cite{Dorogovtsev2008}.
This phenomenon is due to the presence of hubs, \emph{i.e.}, the nodes
with very large degree (diverging with the network size $n$). 
For any value of $\alpha>0$,  the payoffs $\pi_C$ of cooperating 
for the largest hubs will become positive for sufficiently large values of 
$n$. Those hubs then spread the cooperative strategy to the 
rest of the network.

Numerical simulations of PI dynamics on scale-free networks
confirm only in part the theoretical picture (Figure \ref{fig.SC_PI}).
We observe a continuous transition
between a fully defective Nash equilibrium, for small values of $\alpha$ 
and final states with intermediate cooperation (which are not Nash 
equilibria) for $\alpha > \alpha_T$. As $\alpha$ keeps increasing,
cooperation becomes the stable strategy for an increasing number of
nodes, and the system heads smoothly
towards full cooperation---which in this case becomes a stable Nash
equilibrium starting from $\alpha=c/k_{min}$ (which is finite as
$k_{min} > 0$ in the configuration model).
The transition point $\alpha_T$ tends to zero as the system size 
grows.
Hence HMF theory predicts correctly that the fully defective state
disappears in the large size limit (a phenomenon not captured by MF 
or by the one-shot results in \cite{Galeotti2010}), 
but it fails in predicting that the transition is continuous.

\subsubsection*{Best Response}

For coordination games under BR, the approximate MF calculations
predict again the existence of a critical value $\alpha_c = c/(m
\rho_0)$ such that for $\alpha \ll \alpha_c$ the only Nash equilibrium
is full defection, whereas, for $\alpha \gg \alpha_c$ the final state
exhibits a large level $\rho^*$ of cooperation. Note that here
any Nash equilibrium features players with $k<c/\alpha$ being
defectors by construction; therefore, the fully cooperative Nash
equilibrium is achieved for $\alpha \gg \alpha_c$ but only in networks with $k_{min}>1$.

\begin{figure}[h!]
\begin{minipage}[b]{0.49\textwidth}
\centering
\includegraphics[width=0.975\textwidth]{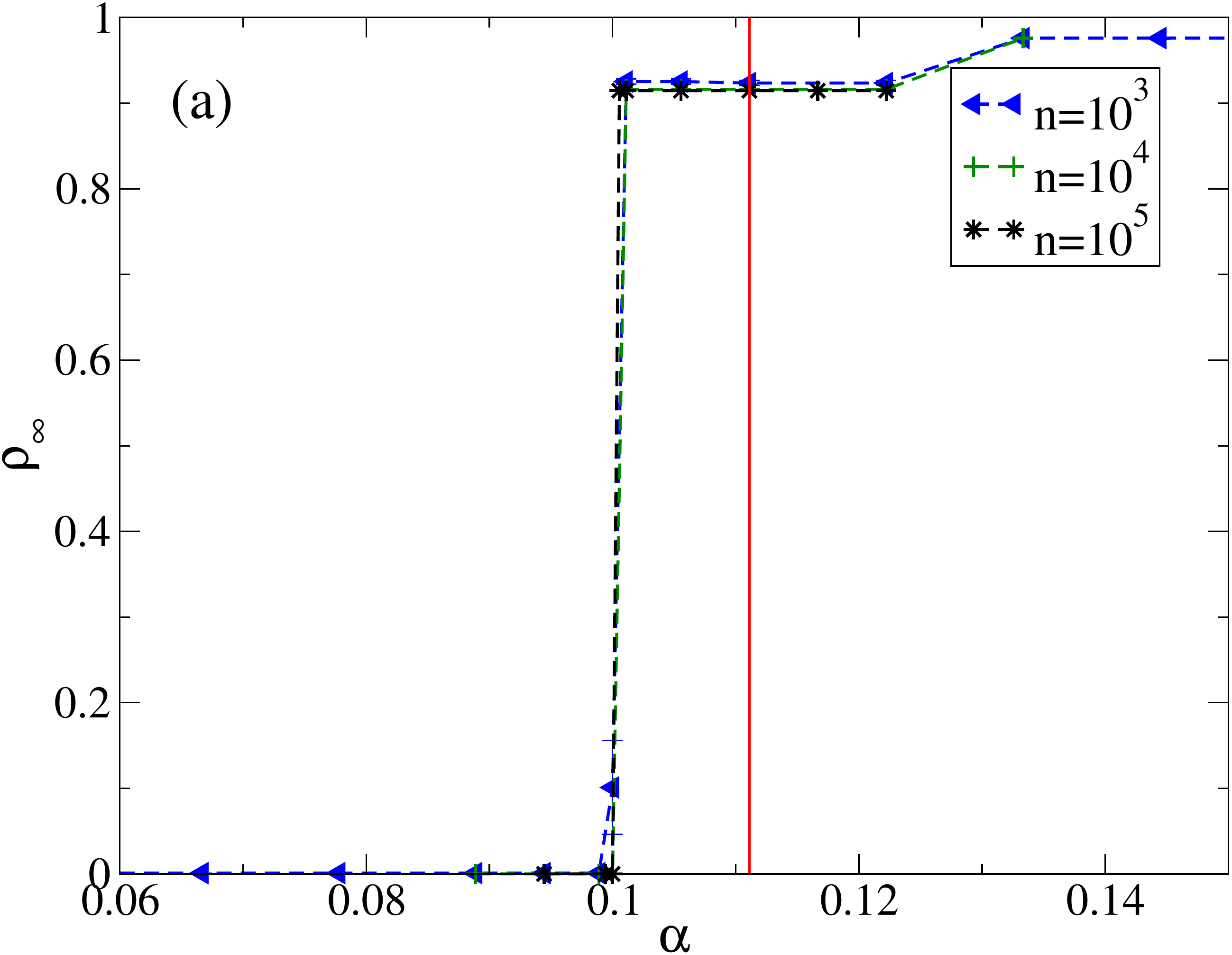}
\end{minipage}
\begin{minipage}[b]{0.49\textwidth}
\centering
\includegraphics[width=0.975\textwidth]{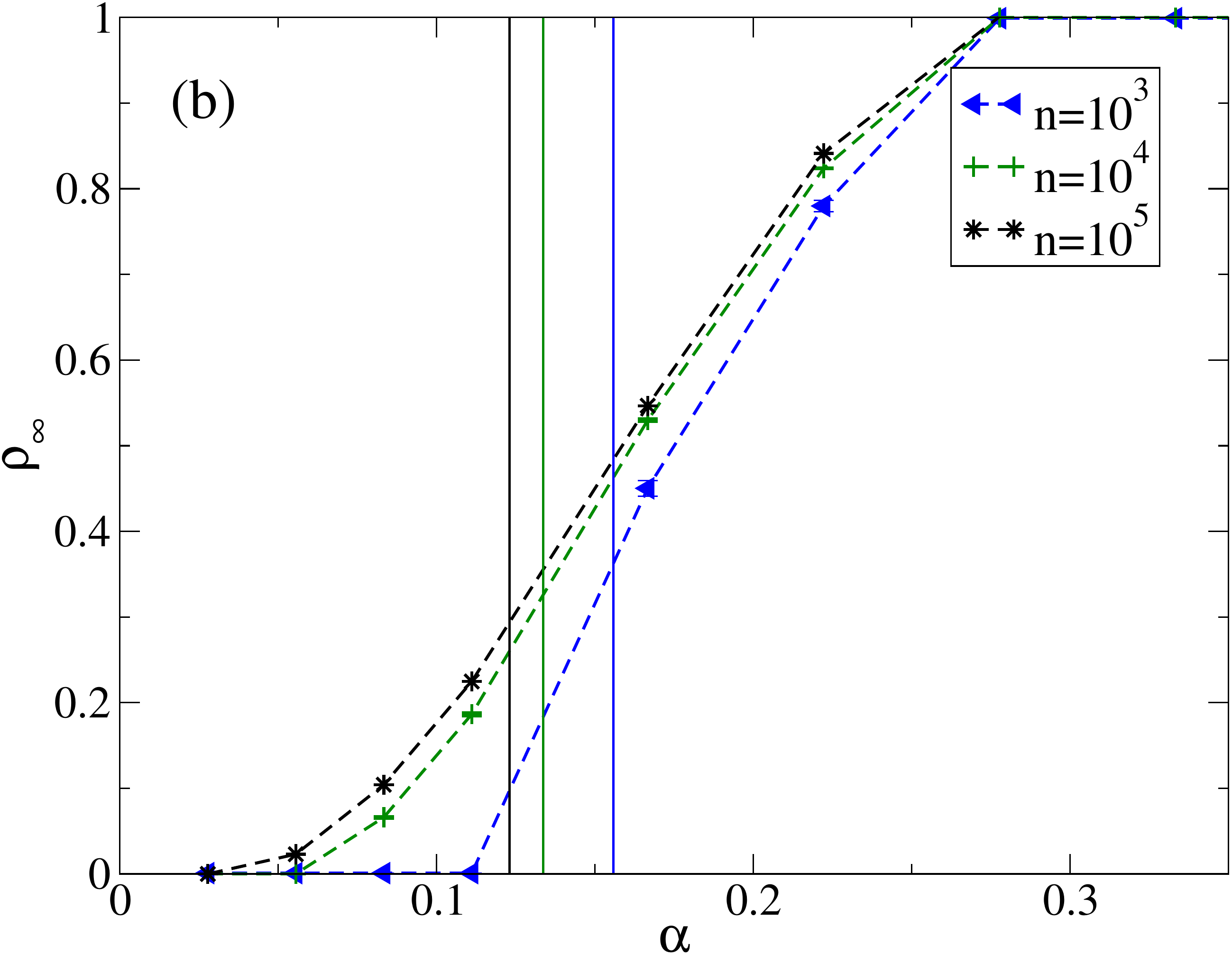}
\end{minipage}
\centering
\includegraphics[width=0.65\textwidth]{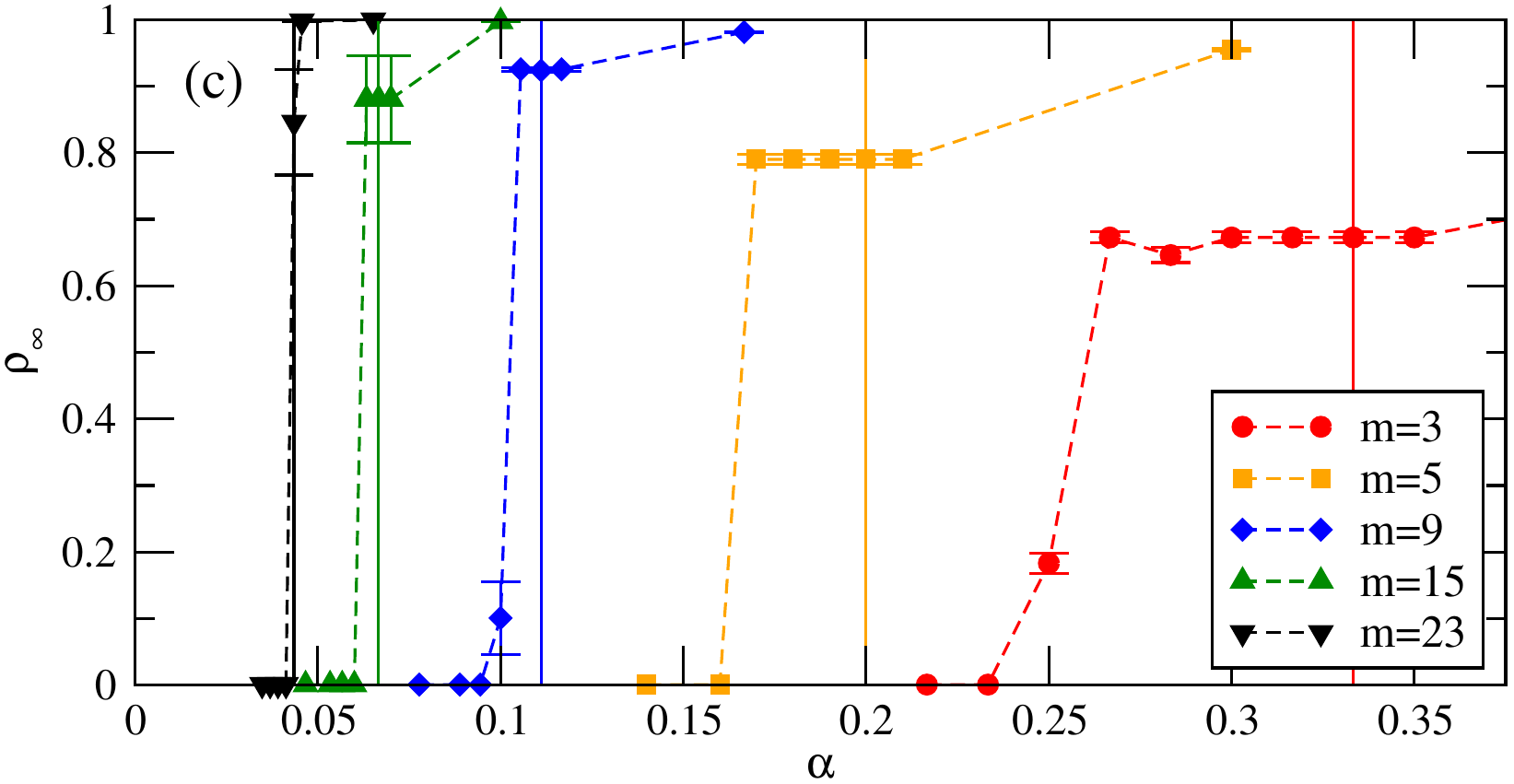}
\caption{Coordination games with Best Response. 
The vertical solid lines identify the values of $\alpha_c$. 
Erd\"{o}s-R\'{e}nyi random graphs: 
(a) stationary  cooperation levels $\rho_\infty$ vs $\alpha$ for $m=9$ and various $n$;  
(c) stationary cooperation levels $\rho_\infty$ vs $\alpha$ for $n=10^3$ and various $m$. 
Note that as $m$ decreases, the amount of nodes with $k<c/\alpha$ increases; 
as these players are defectors by construction,
$\rho_\infty\nrightarrow 1$ even for $\alpha\rightarrow\infty$.
Scale-free networks: 
(b) stationary cooperation levels $\rho_\infty$ vs $\alpha$ for $m=9$ and various $n$.}\label{fig.SC_BR}
\end{figure}

The behavior found in numerical simulations is similar to the case of PI (Figure~\ref{fig.SC_BR}).
On Erd\"{o}s-R\'{e}nyi random graphs  
a discontinuous transition is found (players have
similar degrees, and thus, for a given $\rho$, they become cooperators
for similar values of $\alpha$). An important difference is
that many non-trivial Nash equilibria (with intermediate cooperation
levels $\rho^*$) are found just above the transition point.
A continuous transition is found instead for scale-free networks. 
Here players have different degrees,
and for a given $\rho$ each degree class requires its own value of
$\alpha$ to switch to cooperation.

The HMF approach yields a self-consistent equation 
for the equilibrium $\Theta_s$:
\begin{equation}\label{eq.SC_BR_h_4}
\Theta_s=\sum_{k>c/(\alpha\Theta_s)}kP(k)/\bar{k}
\end{equation}
If the network is scale-free with $2<\gamma<3$, $\Theta_s$
represents a stable equilibrium whose dependence on $\alpha$ is of the
form $\Theta_s\sim\alpha^{(\gamma-2)/(3-\gamma)}$, \emph{i.e.}, there
exists a non-vanishing cooperation level $\Theta_s$ no matter how
small the value of $\alpha$. However, if the network is
homogeneous (\emph{e.g.}, $\gamma>3$), $\Theta_s$ becomes unstable and
for $\alpha \rightarrow 0$ the system always falls in the fully
defective Nash equilibria.

\begin{figure}[h!]
\centering
\begin{minipage}[b]{0.525\textwidth}
\centering
\includegraphics[width=\textwidth]{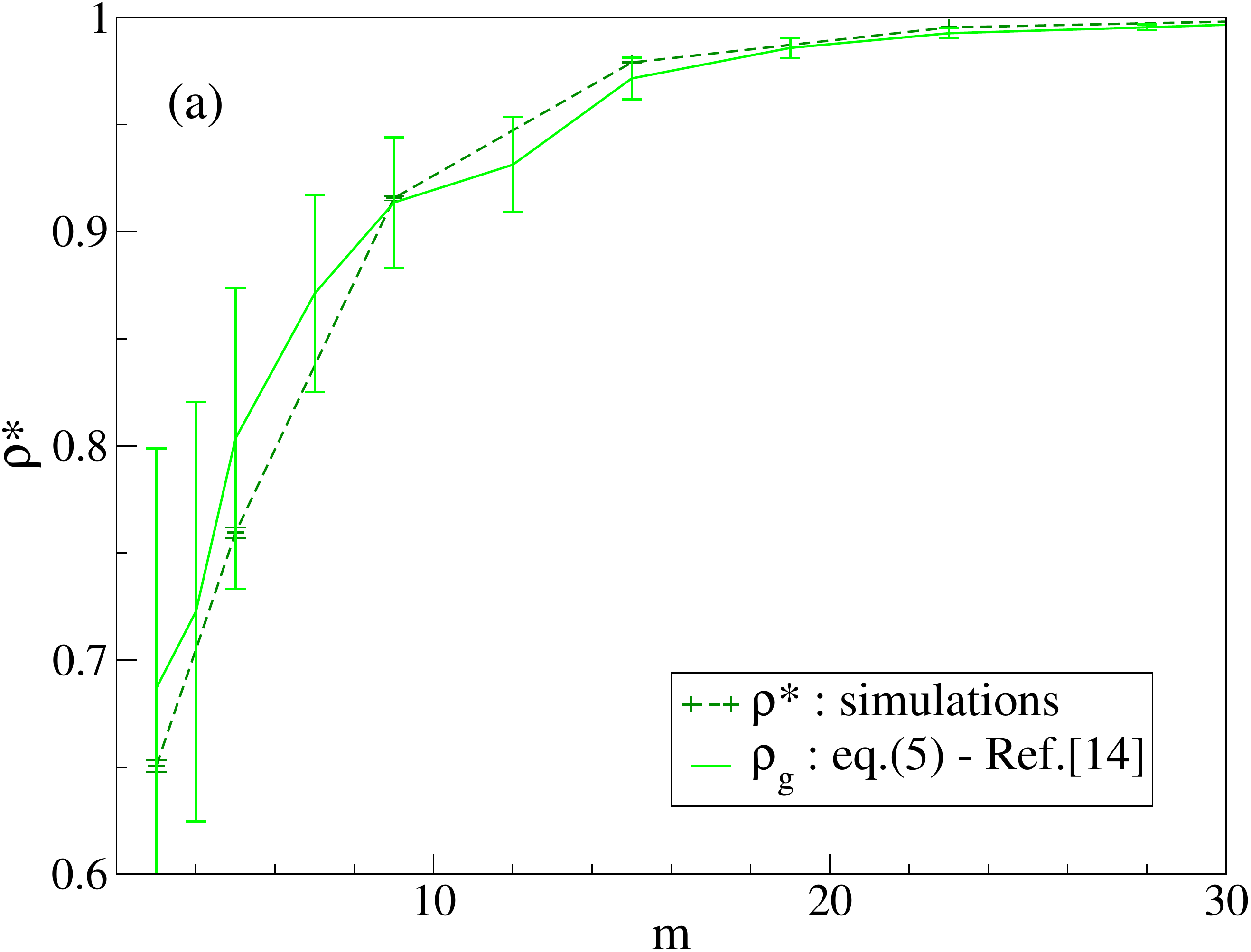}
\end{minipage}
\begin{minipage}[b]{0.45\textwidth}
\centering
\includegraphics[width=\textwidth]{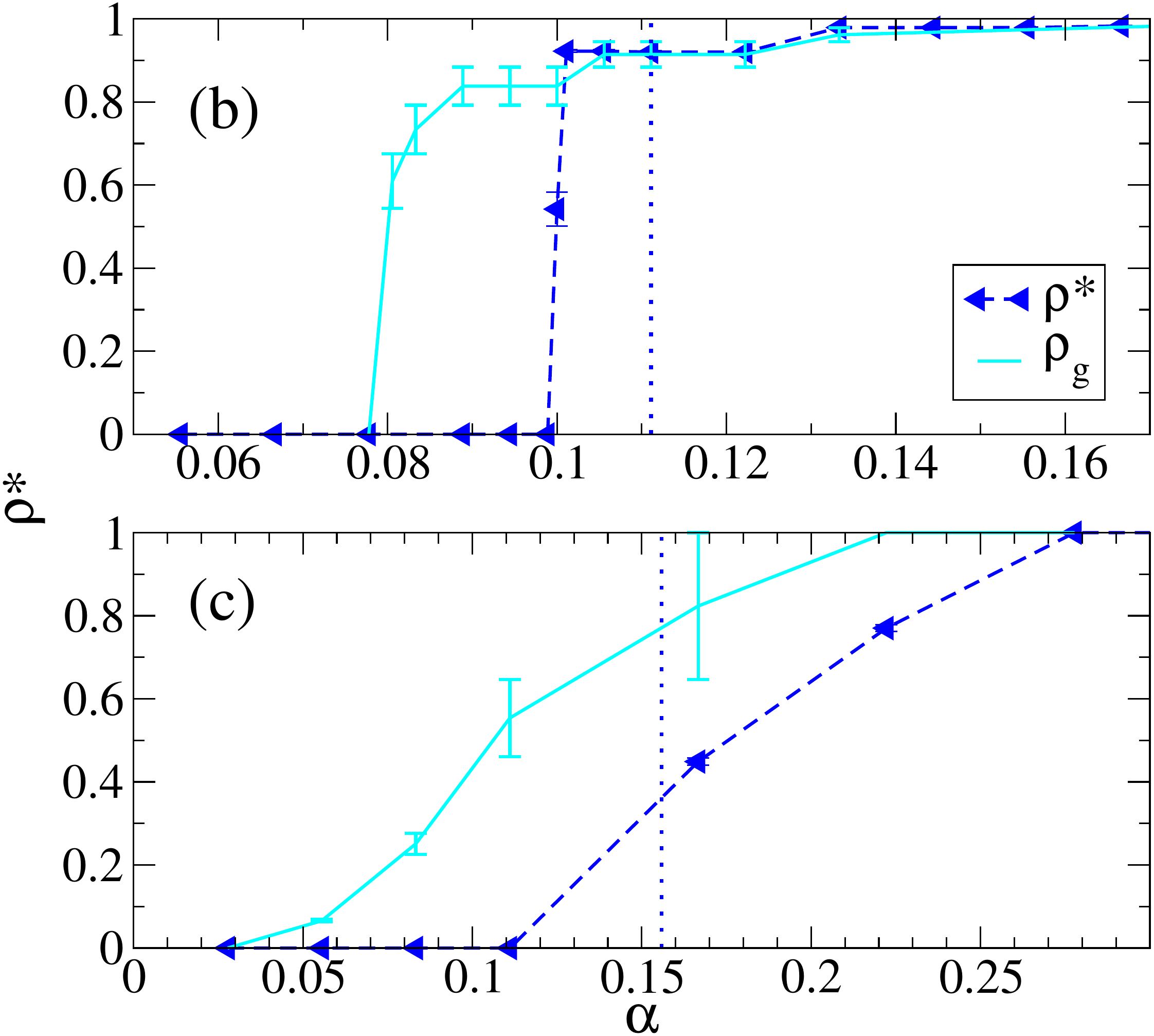}
\end{minipage}
\caption{Coordination games with Best Response: average $\rho^*$ at
  Nash equilibria from simulations and theoretical prediction from
  Eq.(\ref{eq.SC_tau}).  (a) $\rho^*(m)$ for Erd\"{o}s-R\'{e}nyi
  random graphs with $n=10^4$ at $\alpha=\alpha_c$; (b)
  $\rho^*(\alpha)$ for Erd\"{o}s-R\'{e}nyi random graphs with $n=10^3$
  and $m=9$; (c) $\rho^*(\alpha)$ for scale-free networks with
  $n=10^3$ and $m=9$. The vertical dashed lines denote the critical value $\alpha_c$.}
\label{fig.SC_BR_eq}
\end{figure}
\begin{figure}[h!]
\centering
\includegraphics[width=0.85\textwidth]{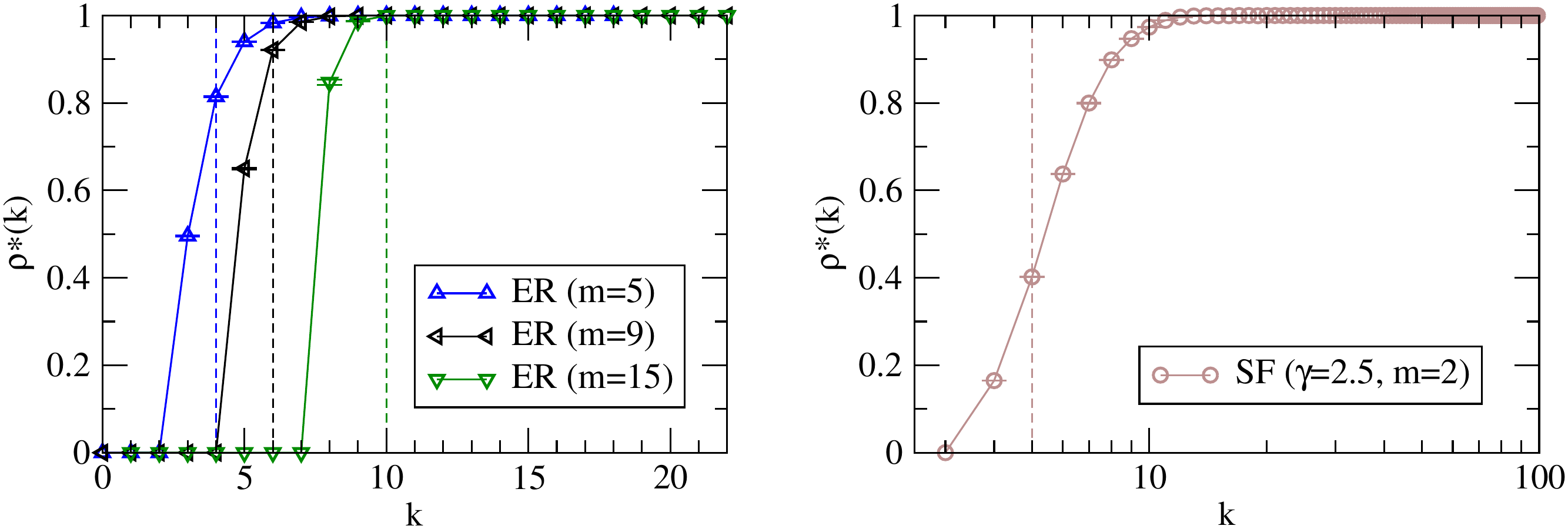}
\caption{Coordination games with Best Response: average $\rho^*(k)$ at
  Nash equilibria at $\alpha=\alpha_c$ for nodes with different
  degrees $k$.  The vertical dashed lines identify the thresholds $\tau$
  form Eq.(\ref{eq.SC_tau}).}\label{fig.SC_BR_degree}
\end{figure}

The existence, in some intervals of $\alpha$ values, of Nash equilibria
with intermediate levels of cooperation calls for the comparison 
between the simulation results and the predictions of Ref.~\cite{Galeotti2010}.
Figure~\ref{fig.SC_BR_eq}(a) shows that, for Erd\"{o}s-R\'{e}nyi random 
graphs at $\alpha=\alpha_c$, the cooperation level of equilibria 
found dynamically lies in the range $\rho_g\in[\rho_{k\ge\tau},\rho_{k>\tau}]$, 
where $\tau$ is given by Eq.(\ref{eq.SC_tau}).
However, we do not find such a good agreement for all values of 
$\alpha$ nor for scale-free networks (Figures \ref{fig.SC_BR_eq}b,c).

Another important prediction of the HMF approach is that
$\rho_k \rightarrow 0$ when $k<c/(\alpha\Theta_s)$, and
$\rho_k \rightarrow 1$ for $k>c/(\alpha\Theta_s)$. 
In this sense, the equilibria predicted by HMF agree qualitatively with
those found in \cite{Galeotti2010}: players' actions show a
non-decreasing dependence on their degrees. 
Indeed, Figure~\ref{fig.SC_BR_degree} shows that the average cooperation 
level $\rho^*(k)$ of Nash equilibria found dynamically in simulations
for $\alpha=\alpha_c$ is generally non-decreasing in $k$, and that a
step-like behavior of $\sigma(k)$, as predicted in~\cite{Galeotti2010}, 
becomes a good approximation for large $n$.
In conclusion, the equilibria
predicted in~\cite{Galeotti2010}, are a good approximation of the
equilibria found dynamically only in the thermodynamic limit 
($n\rightarrow\infty$) and, more importantly, only for a small 
subset of $\alpha$ values.

\subsection{The effect of correlated networks}

In order to study the effects of topologies with degree-degree correlations 
on the behavior of the games, we run simulations with interaction
patterns given by scale-free networks with assortative and disassortative
correlations.
In particular, we consider networks generated using the prescriptions
of the model by Weber and Porto~\cite{Weber2007}, again with the
constraint $k_{max}<\sqrt{n}$ on the largest degree.
The correlation properties of the resulting topologies are dependent on
the model parameter $\beta$.
The average degree of the nearest-neighbors of a node of degree $k$,
$k_{nn}(k):=\sum_j j P(j|k)$, is proportional to $k^\beta$, 
so that for $\beta>0$ neighbors of nodes with large $k$ have large degree
(assortative networks), while for $\beta<0$ nodes with large $k$
have neighbors with small degree and vice versa (disassortative networks).
The uncorrelated case is recovered for $\beta=0$.

\begin{figure}[h!]
\centering
\begin{minipage}[b]{0.525\textwidth}
\centering
\includegraphics[width=\textwidth]{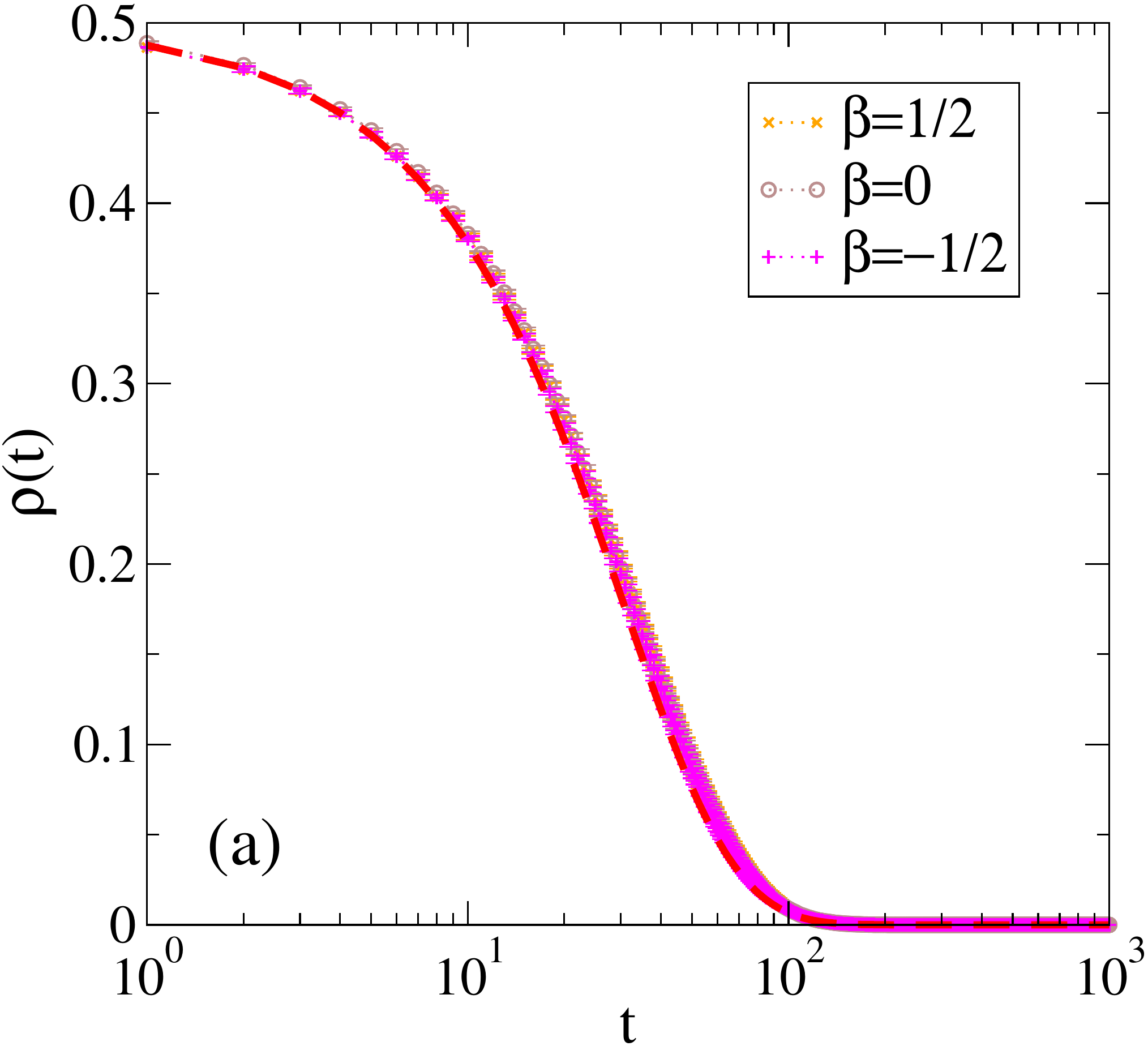}
\end{minipage}
\begin{minipage}[b]{0.45\textwidth}
\centering
\includegraphics[width=\textwidth]{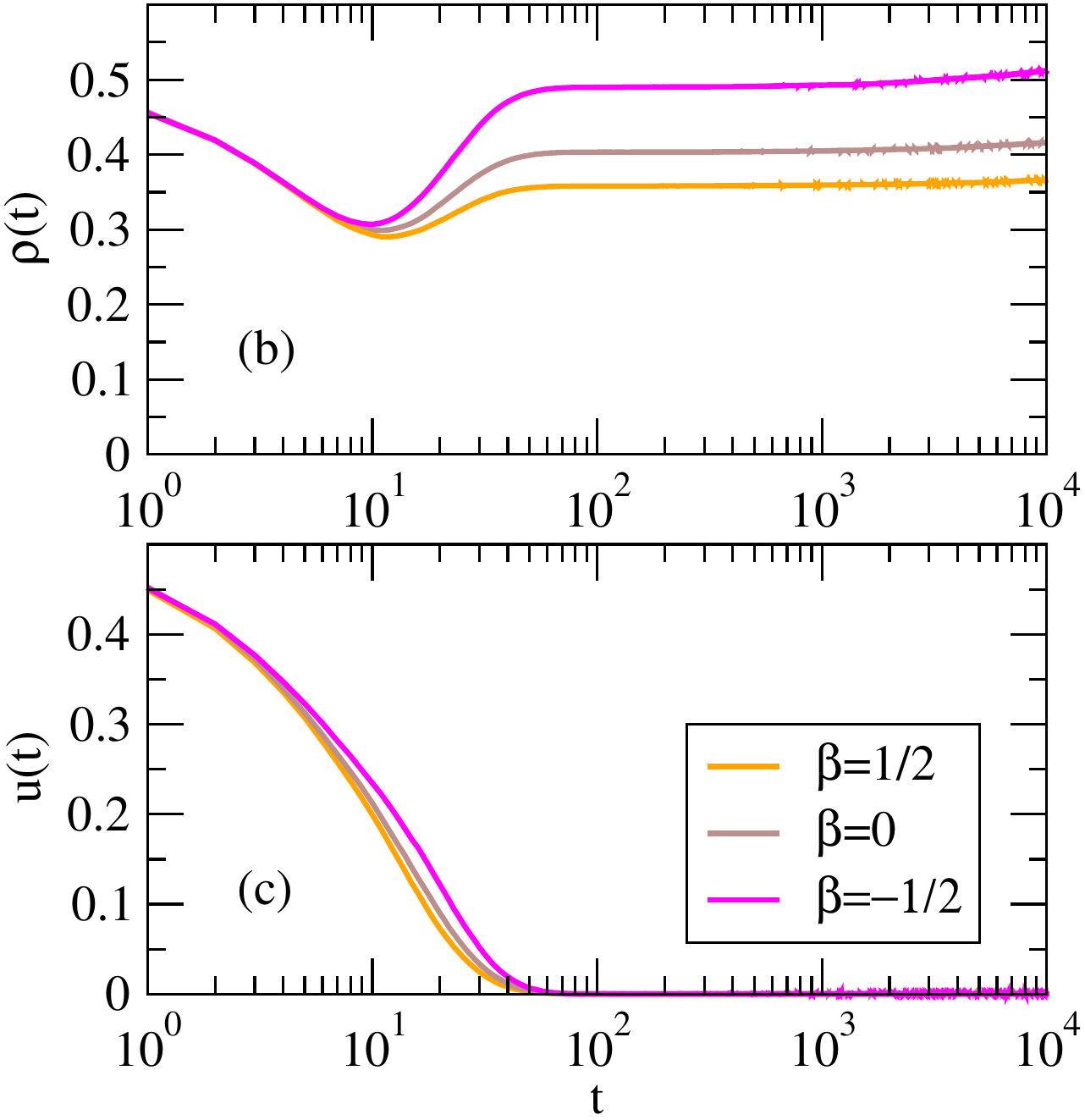}
\end{minipage}
\caption{Best-shot games for correlated networks with $m=9$, $n=10^4$ and various $\beta$. 
Again, these results are independent on the specific value of $n$. 
Proportional Imitation: (a) $\rho(t)$---the red dashed curve being the MF Eq.(\ref{eq.SS_PI_sol}). 
Best Response: (b) $\rho(t)$ and (c) fraction $u(t)$ of unsatisfied players 
(with $u=0$ indicating the falling of the system into a Nash equilibrium).}
\label{fig.SS_wp}
\end{figure}

For the best-shot game (Fig.~\ref{fig.SS_wp}), the simple case of PI dynamics is totally 
unaffected by the presence of correlations: the temporal evolution and the asymptotic state
do not depend at all on $\beta$. 
Instead, when strategies evolve according to BR, the dynamics
leads to Nash equilibria with levels $\rho^*$ of cooperation 
(dominated by the behavior of low-degree nodes) that decrease with $\beta$. 
The reason is that for $\beta<0$ low-degree
nodes are connected to hubs, which are likely to be defectors;
as a consequence they tend to cooperate and $\rho^*$ is higher.
The opposite occurs in assortative networks ($\beta>0$): 
low-degree nodes are connected to each other and less cooperation
is needed.

\begin{figure}[h!]
\centering
\begin{minipage}[b]{0.49\textwidth}
\centering
\includegraphics[width=\textwidth]{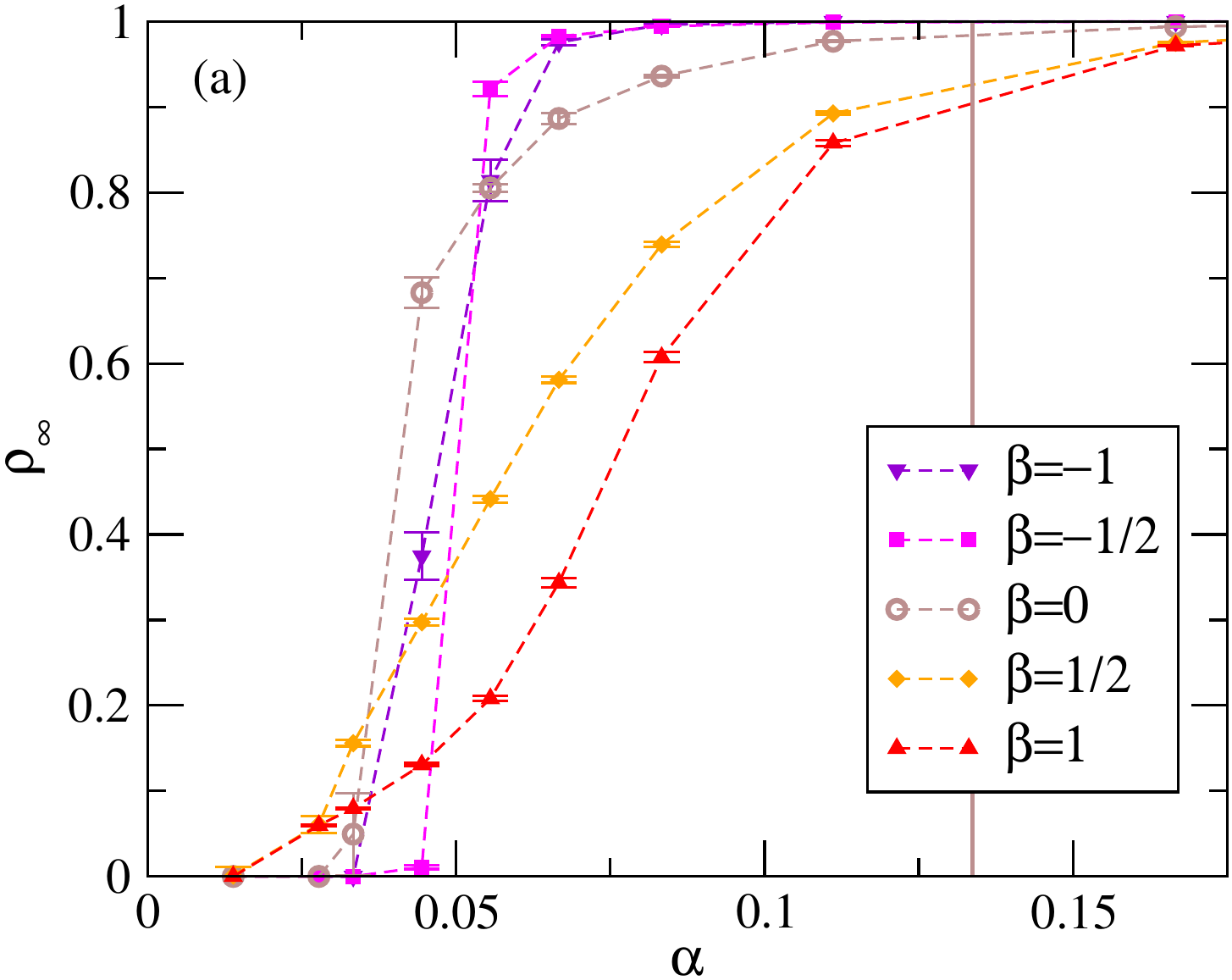}
\end{minipage}
\begin{minipage}[b]{0.49\textwidth}
\centering
\includegraphics[width=\textwidth]{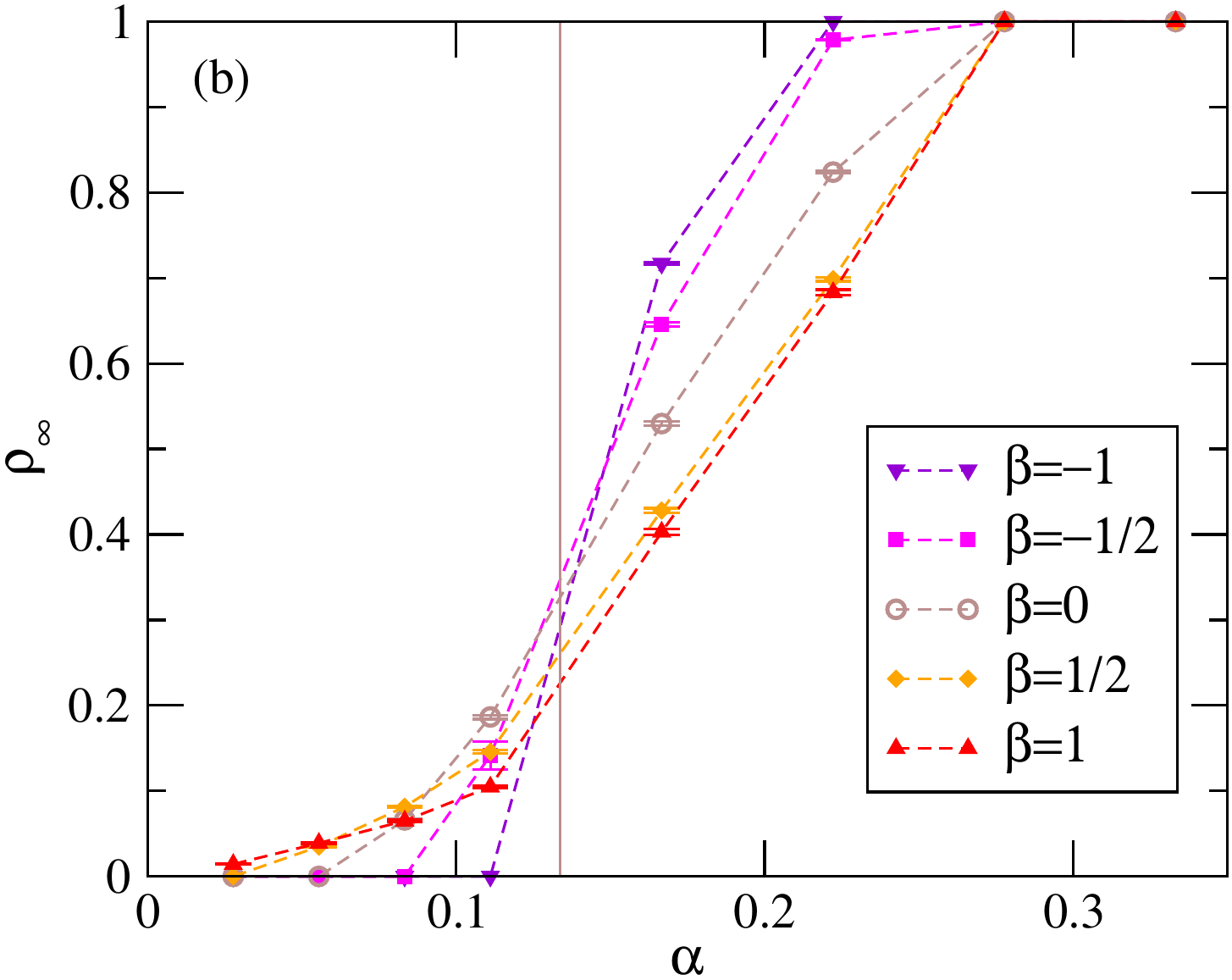}
\end{minipage}
\caption{Coordination games with PI (a) and BR (b) for correlated networks 
with $m=9$, $n=10^4$ and various $\beta$. 
Again, these results are independent on the specific value of $n$. 
The vertical lines identify the values of $\alpha_c$.}
\label{fig.SC_wp}
\end{figure}

For coordination games with PI dynamics,
the effect of assortative (disassortative) correlations is to
make broader (sharper) the transition observed as a function 
of $\alpha$ for finite network size $n$ (Fig.~\ref{fig.SC_wp}).
This can be understood by considering that, in the assortative case,
hubs tend to be connected with each other, so that cooperation
can spread more easily among them and be sustained by
mutual connections: the critical density of stable cooperators
decreases. At the same time, low-degree nodes are connected
only with each other, so that they require larger values of $\alpha$
to sustain cooperation within themselves (as compared to the
uncorrelated case). As a result, the transition becomes broader.
An analogous argument explains the sharper transition for 
dissortative networks. 
In the case of BR dynamics, the same picture applies. 

In conclusion, degree-degree correlations of the interaction pattern
do not change the qualitative features of the behavior of games
on scale-free networks.

\section{Summary and Conclusion}

In this work we have studied by numerical simulations two kinds of
games, namely the best-shot and the coordination games, as
representatives of two wide classes of social interactions (strategic
substitutes and strategic complements, respectively). 
In these games, the welfare of a player depends both on her own actions 
and on the actions taken by her partners; thus, we have considered different
topological structures for the pattern of interactions. 
We have embedded these games into an evolutionary framework, described 
by two types of dynamics, widely employed to model evolving populations: 
Proportional Imitation, in which players imitate more successful 
neighbors, and Best Response, in which players are rational and make 
optimal choices. 
We have performed numerical simulations, determining the attractors 
of the evolution, and characterizing them in term of Nash equilibria.
By comparison with the results of numerical simulations we have assessed
the validity of the mean field approaches in describing such systems, 
and we have also compared our findings with the theoretical
predictions of~\cite{Galeotti2010} about the features of Nash
equilibria in one-shot games under incomplete information.

Generally, we observe that the behavior of the system is highly
influenced by the dynamics employed and by the population structure.
Strategic substitutes under PI dynamics represents the simplest case, 
in which full defection is the only accessible (but non-Nash) 
equilibrium, whatever the underlying topology, in complete agreement 
with the mean field predictions. 
This suggests that the failure to find a Nash equilibrium
arises from the (bounded rational) dynamics and, 
with the benefit of hindsight, it is clear that imitation is not a good
procedure for players to decide in anti-coordination games.  
Such a conclusion is supported by the behavior observed for
strategic substitutes under BR dynamics. 
Indeed, here we observe many stable Nash equilibria, with cooperation level
$\rho^*$ slightly smaller than (but close to) the mean field prediction 
$\rho_c$.
Moreover,  in simulations we observe precisely what is predicted by the
MF theory, namely that $\rho^*$ decreases with increasing network
connectivity and does not depend on the initial conditions, game and simulation parameters,
and system size (which was taken as infinite in the analytical calculations).
Concerning the topology, $\rho^*$ is enhanced in heterogeneous networks 
because of more low-degree nodes who are typically cooperators.  
Additionally, disassortativity has a positive effect on
the cooperation levels, as low-degree nodes are connected to
high-degree nodes who are likely to be defectors, and thus tend to
cooperate even more. Vice versa, assortativity allows more low-degree
nodes to defect.

The picture is far more rich and interesting for strategic
complements, that feature an additional parameter $\alpha$ playing a
key role in determining which equilibria are dynamically accessible.  
Indeed, in the case of PI for Erd\"{o}s-R\'{e}nyi random
graphs we observe two kinds of stationary states: fully
defective Nash equilibria for $\alpha<\alpha_T$, and full cooperation
for $\alpha>\alpha_T$ (that becomes Nash equilibrium only when
$\alpha>c/k_{min}$). Remarkably, $\alpha_T \rightarrow \alpha_c$ for
$n,m \rightarrow \infty$, where $\alpha_c=c/(m\rho_0)$ is the value
predicted by the MF theory. We thus see that imitation is indeed a
good procedure to choose actions in a coordination setup: 
PI does lead to Nash equilibria, and indeed it makes a very precise
prediction: a unique equilibrium that depends on the initial
density. 
If the topology is scale-free, HMF theory
and simulations agree on $\alpha_T$ and $\alpha_c$ going to zero for
$n\rightarrow\infty$, indicating that in these cases cooperation
emerges also when the incentive to cooperate vanishes.  The situation
with BR dynamics is rather similar to the case of PI, with the single
difference that for Erd\"{o}s-R\'{e}nyi random graphs
and $\alpha>\alpha_c$ the stationary state is now a Nash equilibrium,
and again full cooperation is achieved for $\alpha>c/k_{min}$.
Thus we see that, with BR, equilibria with
intermediate values of the density of cooperators are obtained in a range
of initial densities. Compared to the situation with PI, in which we only
find the absorbing states as equilibria, this points to the fact that
more rational players can eventually converge to equilibria with
higher payoffs.
Finally, we have found that, whatever the dynamics, the presence of topological
correlations in the network does not change the global qualitative
picture.

Besides the general features of the dynamics, we have also been able
to study the degree-dependent features of Nash equilibria (when
present) and compare our findings with the theoretical predictions of
\cite{Galeotti2010}, in the which authors consider games played once and
for all but using only partial information on the underlying topology.
Our conclusion is that, while the Nash
equilibria predicted in \cite{Galeotti2010} cannot (always) be
reached, their theory still provides good guidance on the shape and
features of the Nash equilibria which are evolutionarily accessible.

Finally, it is interesting to note that, as in \cite{Roca2009a,Roca2009b}, 
we find that the outcome of the evolution depends on the dynamics 
and the network properties. Remarkably, here we go beyond those 
works as our study of individual behaviors allows us to make 
a connection with the results obtained in a more traditional economic framework.

\end{document}